\documentclass[%
 reprint,
 amsmath,amssymb,
prl,
]{revtex4-2}

\usepackage{graphicx}
\usepackage{dcolumn}
\usepackage{bm}

\begin{document}


\title{Trapping and sympathetic cooling of conformationally selected molecular ions}

\author{Lei Xu}
\author{Jutta Toscano}
\author{Stefan Willitsch}
 \email{stefan.willitsch@unibas.ch}
\affiliation{
 Department of Chemistry, University of Basel, Klingelbergstrasse 80, Basel 4056, Switzerland}

\date{\today}

\begin{abstract}
We report the generation, trapping and sympathetic cooling of individual conformers of molecular ions with the example of {\it cis-} and {\it trans-}{\it meta}-aminostyrene. Following conformationally selective photoionization, the incorporation of the conformers into a Coulomb crystal of laser-cooled calcium ions was confirmed by fluorescence imaging, mass spectrometry and molecular dynamics simulations. We deduce the molecules to be stable in the trap environment for more than ten minutes. The present results pave the way for the spectroscopy and controlled chemistry of distinct ionic conformers in traps.
\end{abstract}

\maketitle

Conformational isomers, or conformers, are molecules with different spatial orientations of the constituent atoms which interconvert through rotation around a single bond. For complex molecules, conformers often represent the dominant isomers responsible for their different possible shapes. Different conformations can exhibit markedly distinct physical and chemical properties, e.g., spectral fingerprints \cite{Stearns2007}, photodissociation dynamics \cite{Park2002,Kim2007, Dian2004} and reactivities \cite{Chang2013,Taatjes2013,Kilaj2021,Kilaj2023}. However, chemical studies unraveling characteristic conformational effects are scarce due to the challenges associated with their selective preparation. This requires specific conformers either to be separated from a mixture \cite{Filsinger2009} or selectively generated \cite{Park2002} in an environment which prevents their isomerization over experimental timescales. The spatial separation of neutral conformers based on their effective dipole moments has recently enabled the characterization of their specific reactivities in a range of reactions \cite{Chang2013,Kilaj2021,Kilaj2023}. However, to date, there is still a lack of methods addressing individual conformers of molecular ions. 

Coulomb crystals represent an attractive environment for the manipulation and control of ionic species \cite{Drewsen2002,Willitsch2012}. These are ordered ensembles of ions held at millikelvin temperatures in radiofrequency (RF) traps in ultrahigh vacuum. Typical experiments employ a laser-coolable atomic species which can sympathetically cool molecular ions into the crystal \cite{molhave00a, Ostendorf2006, Hoejbjerre2008}. Molecular Coulomb crystals have been employed in chemical experiments focussing on state-specific \cite{Tong2010,Tong2012,Hall2012,Doerfler2019} and isomer- and isotope-dependent \cite{Schmid2020,Krohn2021,Yang2021,Staanum2008,Petralia2020,Tsikritea2021} reactivities. These studies yielded new insights into reaction mechanisms and, importantly, proved that Coulomb crystals represent a suitable environment in which individual structural isomers \cite{hojbjerre08a, Yang2021} and even isolated quantum states \cite{tong11a, Tong2012, sinhal20a} can be preserved.  

\begin{figure}[t]
\includegraphics{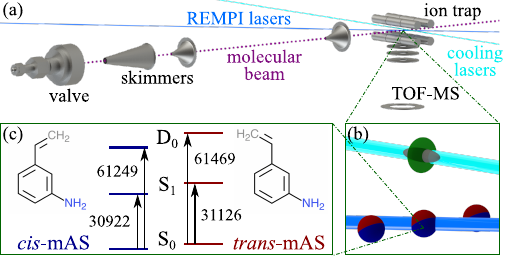}
\caption{\label{fig:Exp} (a) Schematic of the experiment. (b) Separation between the trapping and loading regions: the REMPI lasers intersected the pulsed molecular beam (red and blue spheres) below the center of the ion trap containing a Coulomb crystal (Ca$^+$ in grey, mAS$^+$ in green). (c) Energy level diagrams for the {\it cis} (blue) and {\it trans} (red) conformers of mAS (in cm$^{-1}$). See text for details. 
}
\end{figure}

The present study aims at the control and preservation of ionic conformers within a Coulomb-crystal environment. Selective preparation of the {\it cis} and {\it trans} conformers of the prototypical organic {\it meta}-aminostyrene (mAS) ion (Fig.~\ref{fig:Exp}(c)) was achieved by resonantly ionizing individual conformations of the neutral slightly above their relevant ionization thresholds inside an ion trap containing a Coulomb crystal of laser-cooled calcium ions. The sympathetic cooling of the conformationally selected ions into the crystal was observed by imaging, probed by time-of-flight mass spectrometry, elucidated using Molecular Dynamics (MD) simulations and verified by altering the photoionization scheme. Trapping times exceeding 15 minutes were achieved without significant fragmentation of the mAS$^+$ conformers. Quantum-chemical calculations indicate that the conformers are stable in the Coulomb crystal environment. The present methodology renders possible the full conformational control of ion--molecule reactions and the spectroscopy of individual trapped ionic conformers.   

The details of the experiment have been described previously~\cite{Roesch2016,Kilaj2018}. Briefly, the setup consisted of a molecular-beam machine combined with a linear-quadrupole RF ion trap coupled to a time-of-flight mass spectrometer (TOF-MS) (Fig.~\ref{fig:Exp}(a)). {\it Meta}-aminostyrene (Sigma-Aldrich, 97\% purity) was heated to 80$^\circ$C in a sample holder, seeded in neon gas at 2~bar and supersonically expanded into the beam machine through a pulsed gas valve at 20~Hz repetition rate. The resulting molecular beam contained a mixture of both conformers \cite{Dong2014}. Generation of either the {\it cis} or {\it trans} conformer of mAS$^+$ was achieved by selective resonance-enhanced multiphoton photoionization (REMPI) \emph{via} the first electronically excited states (S$_1$) of the neutral (Fig. \ref{fig:Exp}(c)) using one or two Nd:YAG-pumped dye lasers, respectively. REMPI spectra were recorded by operating the ion trap as a Wiley–McLaren TOF-MS repelling molecules ionized from the beam towards a detector \cite{Roesch2016}. 

For the trapping experiments, the trap was operated at RF frequency $\Omega_{\text{RF}} = 2\pi\times3.29~\text{MHz}$ with amplitude $V_{\text{pp}} = 720~\text{V}$ and a static voltage $V_{\text{DC}} \approx 7~\text{V}$ applied to the endcaps. A camera was used to observe resonance-fluorescence of the laser-cooled calcium ions. Doppler laser cooling of Ca$^+$ was achieved by scattering photons on the $(4s)~^2S_{1/2} \leftrightarrow (4p)~^2P_{1/2} \leftrightarrow (3d)~^2D_{3/2}$ optical cycling transitions using diode laser beams at 397~nm and 866~nm \cite{Willitsch2012}.

Ionization of mAS was performed by focusing the REMPI laser(s) with a cylindrical lens ($f=400~\text{mm}$) to a beam size of $\approx10~\text{mm}\times0.3~\text{mm}$ into the supersonic expansion for 2 minutes. The lasers intersected the molecular beam $\approx1.4~\text{mm}$ below the trap center to separate the ionization and trapping regions, thus avoiding fragmentation of the ions through exposure to the REMPI laser beams. This also prevented collisions between the trapped ions and the pulsed flux of incoming neutral mAS molecules (Fig.~\ref{fig:Exp}(b)). To characterize the ions in the trap, TOF mass spectra were collected by ejecting all trapped ions towards the TOF-MS. 

\begin{figure}[t]
\includegraphics{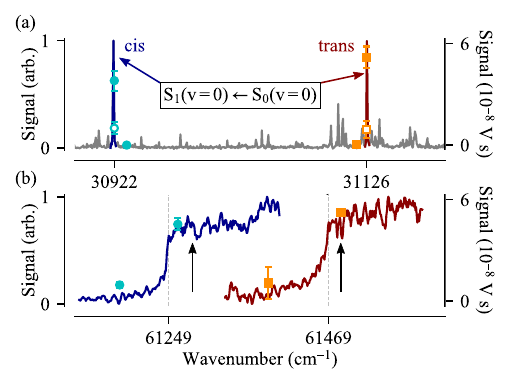}
\caption{\label{fig:REMPI} Photoionization of \emph{meta-}aminostyrene (mAS) conformers: (a) (1~+~1)-photon REMPI spectrum with the S$_1$~($v$~=~0)~$\leftarrow{}$~S$_0$~($v$~=~0) transitions of the two conformers highlighted. Symbols: Integrated mAS$^+$ ion signal at different REMPI lasers frequencies for the {\it cis} (cyan circles) and {\it trans} (orange squares) conformers after sympathetic cooling and ejection of the Coulomb crystal into the TOF-MS. (b) (1~+~1$'$) photoionization spectra for the \emph{cis} (blue) and \emph{trans} (red) conformer with the excitation laser wavenumber fixed to the transitions indicated in (a).  Symbols: Integrated mAS$^+$ ion signal at different ionization laser wavenumbers after sympathetic cooling and ejection of the Coulomb crystal into the TOF-MS. Error bars represent the standard deviation of three measurements. See text for details.}
\end{figure}

The two-photon resonance-enhanced ionization of mAS from the neutral (S$_0$) to the ionic  (D$_0$) electronic ground state \emph{via} the first electronically excited state (S$_1$) of the neutral (Fig. \ref{fig:Exp}(c)) was achieved with either a one-color [(1~+~1)-photon] or a two-color [(1~+~1$'$)-photon] scheme, as demonstrated in Ref.~\cite{Dong2014}. Fig.~\ref{fig:REMPI}(a) shows the one-color REMPI spectrum recorded by monitoring the yield of mAS$^+$ photoionized from the molecular beam as a function of the laser wavenumber. The origin bands, i.e., the vibrationless $(v=0)$ S$_1$~~$\leftarrow{}$~S$_0$~ transitions, of {\it cis}- and {\it trans}-mAS were identified as the excitations marked blue and red in Fig.~\ref{fig:REMPI}(a) based on the assignments of Ref.~\cite{Dong2014} and the photoionization experiments reported below. Other features in the spectrum correspond to transitions to excited vibrational levels in the S$_1$ state of both conformers \cite{Dong2014}. From the spectrum, the resonance frequencies for the origin bands were determined to be 30922~cm$^{-1}$ and 31126~cm$^{-1}$ for {\it cis}- and {\it trans}-mAS, respectively. These values are in agreement with the results of Ref.~\cite{Dong2014}, with a $\approx$15~cm$^{-1}$ discrepancy probably stemming from differences in laser calibration. 

Fig.~\ref{fig:REMPI}(b) displays conformationally selected photoionization (PI) spectra, i.e., the photoion yield as a function of the resonant (1~+~1$'$) two-photon energy, for the \emph{cis} and \emph{trans} species in the blue and red traces, respectively. The PI spectra were taken {\it via} excitation of the S$_1~(v=0)$ state of each conformer, i.e., the transitions marked blue and red in Fig.~\ref{fig:REMPI}(a). From the different onsets of the ionization signals, it can be seen that both conformers have distinct ionization thresholds and can clearly be distinguished in ionization. The ionization energies corresponding to the transitions (D$_0$~($v$~=~0)~$\leftarrow{}$~S$_0$~($v$~=~0)) were estimated from the steep slope of the ion signal in the PI spectra to be 61249~cm$^{-1}$ and 61469~cm$^{-1}$ for {\it cis-} and {\it trans-}mAS, respectively (vertical dashed lines in Fig.~\ref{fig:REMPI}(b)). These estimations are in reasonable agreement, within $\approx$25~cm$^{-1}$, with the previously reported values \cite{Dong2014}.  

\begin{figure}[t]
\includegraphics{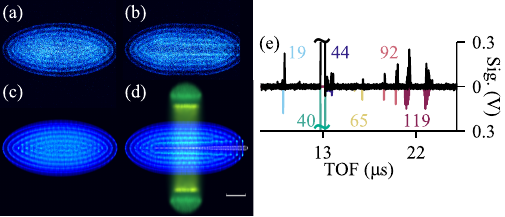}
\caption{\label{fig:Loading} Experimental (a, b) and simulated (c, d) Coulomb crystal images before (a, c) and after (b, d) loading \emph{trans-}mAS$^+$. 
Color code: Ca$^+$ ($m/z = 40$) - blue, mAS$^+$ ($m/z = 119$) - green, C$_6$H$_4$NH$_2^+$ ($m/z = 92$) - yellow, $\text{H}_3\text{O}^+$ ($m/z = 19$) - white. The grey scale bar represents $100~\mu\text{m}$. (e) Experimental (top trace) and simulated (bottom inverted trace) time-of-flight mass spectra of Coulomb crystals after loading. Numbers are atomic mass units of the species.}
\end{figure}

Subsequently, the \emph{cis} and \emph{trans} conformers of the mAS ion were selectively generated in the ion trap by conformer-selective (1~+~1$'$) REMPI slightly above the ionization threshold of the relevant species (black arrows in Fig.~\ref{fig:REMPI}(b)). The generation of Coulomb crystals and sympathetic cooling of the mAS$^+$ conformers was monitored by imaging the laser-cooled calcium ions. Starting from an initially pure Ca$^+$ Coulomb crystal (Fig.~\ref{fig:Loading}(a)), the sympathetic cooling of non-fluorescing molecular ions became visible as dark regions in the center and at the extremities of the crystals in the images (Fig.~\ref{fig:Loading}(b)). 

The various species present in the crystals were identified and quantified by mass spectrometry. Fig.~\ref{fig:Loading}(e) shows a TOF mass spectrum averaged over three typical Coulomb crystals after the ionization of {\it trans-}mAS displaying a number of different mass species. For the interpretation of the spectra, MD simulations of the trajectories of the ejected ions were performed using SIMION \cite{Dahl2000}. A simulated TOF mass spectrum is displayed in the lower inverted trace in Fig.~\ref{fig:Loading}(e). The double peaks observed are due to the radial separation of ions heavier than Ca$^+$ in the trap, as discussed in Ref.~\cite{Roesch2016}. From comparing with experiments, it could be established that besides the parent mAS$^+$ ($\text{C}_8\text{H}_9\text{N}^+$, $m/z$~=~119), the crystals also contained $\text{C}_6\text{H}_4\text{NH}_2^+$ ($m/z$~=~92) resulting from bond cleavage between the ring and the vinyl group. The remaining peaks were attributed to impurities: $^{44}\text{Ca}^+$ ($m/z$~=~44) present in an amount consistent with the 2\% natural abundance of this calcium isotope and $\text{H}_2\text{O}^+$ and $\text{H}_3\text{O}^+$ ($m/z$~=~18,~19) attributed to the ionization and reactions of water impurities in the vacuum and molecular beam.  

The number of mAS$^+$ ions loaded into the crystal in Fig.~\ref{fig:Loading}(b) was inferred to be $\approx400$. The identity and number of ions established by mass spectrometry were then used as input for MD simulations of the crystal images (Fig.~\ref{fig:Loading}(c) and (d)) following Ref.~\cite{Rouse2015}. For the example in Fig.~\ref{fig:Loading}, the initial Coulomb crystal was simulated using 1600 Ca$^+$ ions (Fig.~\ref{fig:Loading}(c)), with the addition of 425 mAS$^+$, 50 $\text{C}_6\text{H}_4\text{NH}_2^+$ and 36 $\text{H}_3\text{O}^+$ ions after loading (Fig.~\ref{fig:Loading}(d)). The MD simulations elucidate the changes in the crystals: ions lighter than Ca$^+$ accumulate at the center of the crystal forming a dark core, whereas heavier species surround it annularly. The non-fluorescing molecular ions were made artificially visible in the simulated image Fig.~\ref{fig:Loading}(d). The simulations yield a secular temperature of $\approx25~\text{mK}$  and $<50~\text{mK}$ for the Ca$^+$ and mAS$^+$ ions, respectively. 

To verify that the ions sympathetically cooled into the Coulomb crystals were indeed conformationally selected, the amount of trapped mAS$^+$ ions was analyzed as a function of the REMPI laser frequencies. This was accomplished by integrating the  mAS$^+$ TOF signal following loading of the crystal at different combinations of excitation and ionization laser frequencies. 

Fixing the ionization laser wavenumber above threshold, an appreciable ion signal was only observed when the laser for the excitation to the S$_1$ state was resonant with a transition in a specific conformer while being negligible when it was slightly off-resonance (filled symbols in Fig.~\ref{fig:REMPI}(a)). With the excitation laser on resonance, blocking the ionization laser dramatically decreased the observed ion signal (empty symbols in Fig.~\ref{fig:REMPI}(a)), demonstrating that the ionization mainly occurred {\it via} (1~+~1$'$) REMPI, with the (1~+~1) contribution only amounting to $\approx10-20\%$. Fixing the excitation-laser wavenumber at the respective resonance for each conformer, appreciable signal was only obtained when the total laser wavenumber was set above the ionization threshold, while below threshold the remaining ion signal level corroborates the contribution from the (1~+~1) REMPI process (cyan and orange symbols in Fig.~\ref{fig:REMPI}(b)). This contribution was carefully minimized by decreasing the intensity of the excitation laser, but attempts to further reduce it resulted in inefficient loading due to poor population of the $\text{S}_1$ state. 

\begin{figure}[t]
\includegraphics{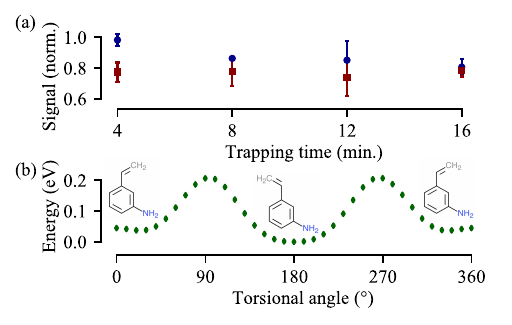}
\caption{\label{fig:Stability} (a) Normalized integrated mAS$^+$ ion signal following (1~+~1$'$) REMPI as a function of the trapping time before ejection into the TOF-MS for {\it cis}- (blue circles) and {\it trans}-mAS$^+$ (red squares). Error bars represent the standard deviation of three measurements. (b) Relative energy of mAS$^+$ as a function of the the interconversion coordinate, i.e., the torsional angle between the vinyl group and the aminophenyl ring plane. Angles of 0\textdegree{} and 360\textdegree{} (180\textdegree{}) correspond to {\it cis}- (\emph{trans-}) mAS$^+$.}
\end{figure}

The stability of {\it cis-/trans-}mAS$^+$ within our experimental environment was verified by examining the number of ions retained in the trap after a variable period of time. The integrated mAS$^+$ signal was not found to decrease significantly over trapping times up to 16 minutes for both conformers (Fig.~\ref{fig:Stability}(a)) indicating negligible loss.

Furthermore, to assess the possibility of isomerization of the conformers in the trap, Density Functional Theory (DFT) and Coupled Cluster (CC) quantum-chemical calculations were carried out to estimate the energy required for {\it cis-trans} interconversion through rotation about the bond between the vinyl and aminophenyl moieties in the molecule. The energy barrier for the interconversion was calculated using GAUSSIAN~16 \cite{Frisch2016}. The energy profile along the interconversion coordinate (Fig. \ref{fig:Stability}(b)) was computed by performing a geometry optimization at fixed values of the torsional angle using DFT at the M06-2X level \cite{Zhao2008} with GD3 dispersion correction \cite{Grimme2010} and the aug-cc-pVTZ basis set \cite{Dunning1989, Kendall1992}. 
For comparison, the isomerization energies were also calculated using a series of alternative DFT methods and the same basis set from the energy difference between optimized structures of {\it cis}- and {\it trans}-mAS$^+$ and the transition state between them. Dispersion corrections were included for all DFT calculations: GD3 for M06-2X, D3BJ for DSD-PBEP86 \cite{Kozuch2011, Kozuch2013} and GD3BJ \cite{Grimme2011} for UB3LYP \cite{Becke1993} and PBE0 \cite{Adamo1999}.
The optimized M06-2X geometries were also used for single-point CC calculations (CCSD and CCSD(T) \cite{Cizek1969, Purvis1982}; aug-cc-pVDZ basis set for both). The results are collected in Table~\ref{tab:Barriers}. 

\begin{table}[t]
\caption{\label{tab:Barriers}%
Calculated isomerization energies (in meV)
}
\begin{ruledtabular}
\begin{tabular}{ldd}
Method & \multicolumn{1}{c}{{\it cis}-to-{\it trans}} & \multicolumn{1}{c}{{\it trans}-to-{\it cis}} \\
\colrule
UB3LYP    & 213                & 265                 \\
PBE0   & 201                & 252                 \\
M06-2X     & 172                & 208                 \\
DSD-PBEP86 & 152                & 194                 \\
CCSD(T)   & 99                 & 134                 \\
CCSD      & 89                 & 121                
\end{tabular}
\end{ruledtabular}
\end{table}

Given the greater stability of the {\it trans-} compared to the {\it cis-}conformer (Fig.~\ref{fig:Stability}(b)), {\it cis}-to-{\it trans} isomerization exhibits slightly lower energy barriers than {\it trans}-to-{\it cis} conversion for all methods considered (Table~\ref{tab:Barriers}). Depending on the method, these barriers range between $89-213~\text{meV}$ and $121-265~\text{meV}$, respectively. These results agree favorably with previous calculations reported in Ref. \cite{Dong2014}. 

The stability of the conformers within the Coulomb crystal environment can be affected by various factors: (i) {\it Laser absorption}: Interaction with laser light can potentially lead to fragmentation or isomerization of the conformers. Once either {\it cis-} or {\it trans-}mAS$^+$ ions were generated, further absorption of the REMPI laser light was prevented by segregating the loading and trapping regions, as outlined above. The cooling laser beams were focused to the crystal center and only weakly overlapped with the trapped mAS$^+$ ions since their $1/e^2$ beam waist diameter ($\approx400~\mu\text{m}$) was smaller than the radial displacement of the mAS$^+$ ions, which accumulated in the range $\approx510-560~\mu\text{m}$ from the trap center according to the MD simulations (Fig.~\ref{fig:Loading}(d)). The lack of interaction between the trapped conformers and the cooling light is substantiated by the negligible fragmentation of the mAS$^+$ ions  (Fig.~\ref{fig:Stability}(a)). (ii) {\it Collisions}: Sympathetic cooling involves elastic collisions between ions which decrease the kinetic energy but do not change the internal energy of the trapped molecular ions \cite{Willitsch2012}. Moreover, separating the molecular beam from the trap center minimized collisions of trapped ions with neutral mAS and neon seed gas. The former would probably have lead to reactions and can be excluded based on the absence of reaction products in the TOF spectra. The latter could have resulted in elastic and inelastic energy transfer, but the collision energy ($\approx80~\text{meV}$) would have been insufficient to cause isomerization of the conformers (Table~\ref{tab:Barriers}) even in the unlikely event of a total energy transfer from collisional to internal energy of the mAS$^+$ ions. Lastly, collisions with background-gas molecules (such as residual H$_2$, H$_2$O, N$_2$ and O$_2$) could also be inelastic but are deemed to be negligible because low-energetic and infrequent in the ultrahigh vacuum environment ($\approx1\times10^{-10}~\text{mbar}$). (iii) {\it Black-body radiation}: Although {\it cis-} and {\it trans-}mAS$^+$ were initially generated in the D$_0$~($v$~=~0) state by threshold ionization, absorption of black-body radiation (BBR) emitted by the room-temperature environment could lead to the population of excited rovibrational states on the time scale of trapping experiments \cite{tong11a, deb14a}. However, as the isomerization barriers (Table \ref{tab:Barriers}) are considerably higher than the thermal energy available in the present case ($k_{\text{B}}T=26~\text{meV}$), interconversion of the conformers through optical pumping by BBR is expected to be slow. 

We thus surmise that the conformation of the sympathetically cooled mAS$^+$ ions is stable over the current experimental timescales in the Coulomb-crystal environment. This can be put into context with previous evidence that even single molecular quantum states can be retained for minutes under these conditions \cite{tong11a, sinhal20a}. A more detailed characterization of the conformational dynamics would require direct spectroscopy of the trapped mAS$^+$ which, however, yet remains to be established. 

In summary, conformationally selected {\it meta-}aminostyrene ions were successfully generated and sympathetically cooled into a Coulomb crystal. This was achieved by spatially separating the loading and trapping regions which proved crucial to minimize fragmentation of the trapped conformationally selected ions. The methodology demonstrated here is widely applicable and can straightforwardly be extended to the trapping and cooling of other large molecules whose conformational isomers can be selectively ionized. The number of trapped mAS$^+$ conformers realized in the present work and their stability to the experimental environment is sufficient for future reaction and spectroscopic studies with conformationally selected trapped ions. In particular, combined with the capability to select the conformation of the neutral counterpart \cite{Chang2013}, fully conformationally controlled ion-molecule reaction studies now come within reach. 

\begin{acknowledgments}
We acknowledge technical support from Dr. A. Johnson, G. Holderried, G. Martin and Ph. Kn\"opfel. This work was funded by the Swiss National Science Foundation, grant nr. IZCOZ0\_189907, and the University of Basel.
\end{acknowledgments}

%


\end{document}